\documentclass{article}
\usepackage{spconf,amsmath,graphicx}
\usepackage{color,soul}
\usepackage{hyperref}
\hypersetup{bookmarksnumbered,unicode}
\usepackage{booktabs}
\usepackage{caption}
\usepackage{subcaption}
\usepackage{balance}
\usepackage{threeparttable}%
\usepackage{multirow}
\usepackage{booktabs}
\usepackage[export]{adjustbox}
\usepackage[ruled, linesnumbered]{algorithm2e}

\newcommand{\ie}{i.\,e.,\ }

\title{Exploring Automatic COVID-19 Diagnosis \\ via voice and symptoms from Crowdsourced Data}
\name{Jing Han, Chlo{\"e} Brown$^\ast$, Jagmohan Chauhan$^\ast$, Andreas Grammenos$^\ast$
\thanks{$^\ast$ Ordered alphabetically, equal contribution. This work was supported by ERC Project 833296 (EAR).}}{
Apinan Hasthanasombat$^\ast$, Dimitris Spathis$^\ast$, Tong Xia$^\ast$, Pietro Cicuta, Cecilia Mascolo}
\address{University of Cambridge\\ jh2298@cam.ac.uk}
\begin{document}
\ninept
\maketitle
\begin{abstract}
The development of fast and accurate screening tools, which could facilitate testing and prevent more costly clinical tests, is key to the current pandemic of COVID-19.
In this context, some initial work shows promise in detecting diagnostic signals of COVID-19 from audio sounds.
In this paper, we propose a voice-based framework to automatically detect individuals who have tested positive for COVID-19. We evaluate the performance of the proposed framework on a subset of data crowdsourced from our app, 
containing 828 samples from 343 participants. 
By combining voice signals and reported symptoms, an AUC of $0.79$ has been attained, with a sensitivity of $0.68$ and a specificity of $0.82$.
We hope that this study opens the door to rapid, low-cost, and convenient pre-screening tools to automatically detect the disease. 

\end{abstract}
\vspace{-.1cm}
\section{Introduction}
\label{sec:intro}
\vspace{-.1cm}
On 11 March 2020, the World Health Organisation announced the COVID-19 outbreak as a global pandemic. At the time of writing this paper, more than 37 million confirmed COVID-19 cases and one million deaths globally have been reported. 
Nowadays, in addition to developing drugs and vaccines for treatment and protection~\cite{lurie2020developing, sanders2020pharmacologic}, scientists and researchers are also investigating primary screening tools that ideally should be accurate, cost-effective, rapid, and meanwhile easily accessible to the mass at large.

Amongst 
the efforts towards rapid screening~\cite{mei2020artificial, Brown20-Exploring}, audio-based diagnosis appears promising, mainly due to its non-invasive and ubiquitous  character, which would allow for individual pre-screening `anywhere’, `anytime’, in real-time, and available 
to `anyone’~\cite{schuller2020covid}. Many applications have been developed for monitoring health and wellbeing in recent times via intelligent speech and sound analysis~\cite{ren2015fine, zhang2019snore, huang2019speech}.

COVID-19 is an infectious disease, and most people infected with the COVID-19 experience mild to moderate respiratory illness~\cite{Guan20clinical}.
More specifically, on the one hand, COVID-19 symptoms vary widely, such as cough, dyspnea, fever, headache, loss of taste or smell, and sore throat~\cite{sudre2020symptom}. On the other hand, however, many symptoms are associated with and hence can be recognised via speech and sound analysis. Such symptoms include shortness of breath, dry or wet cough, dysphonia, fatigue, to name but a few. As a consequence, most recently several research works have been published, aiming at providing sound-based automatic diagnostic solutions~\cite{Brown20-Exploring,imran2020ai4covid, bagad2020cough}.

In this paper, we propose machine learning models for voice-based COVID-19 diagnosis.
More specifically, we analyse a subset of data from 343 participants crowdsourced via our app, and show the discriminatory power of voice for the diagnosis. We demonstrate how voice can be used as signal to distinguish symptomatic positive tested individuals, from non-COVID-19 (tested) individuals, who also have developed symptoms akin to COVID-19. 
We further show performance improvement by combining sounds and symptoms for the diagnosis, yielding a specificity of $0.82$ and an AUC of $0.79$.
\vspace{-.1cm}
\vspace{-.3cm}
\section{Related Work}

\label{sec:relatedwork}
With the advent of COVID-19, researchers have started to explore if respiratory sounds could be diagnostic~\cite{schuller2020covid}. For instance, in~\cite{Brown20-Exploring}, breathing and cough sounds have been targeted and researchers demonstrate that COVID-19 individuals are distinguishable from  healthy controls as well as asthmatic patients. In~\cite{ankit2020pay}, an interpretable COVID-19 diagnosis framework has been devised to distinguish COVID-19 cough from other types of cough. Likewise, in~\cite{bagad2020cough}, a detectable COVID-19 signature has been found from cough sounds and can help increase the testing capacity.

However, none of the aforementioned efforts have analysed the potential of voice.
Recently, the feasibility for COVID19 screening using voice has been introduced in~\cite{pinkas2020sars}. Similarly, in~\cite{asiaee2020voice}, significant differences in several voice characteristics are observed between COVID-19 patients and healthy controls.
Moreover, in~\cite{Han20-An}, speech recordings from hospitalised COVID-19 patients are analysed to categorise their health state of patients. 
Our work differs from these works, as we utilise an entirely crowdsourced dataset, for which we have to deal with the complexity of the data such as recordings in different languages and varied environmental noises. 
Furthermore, we jointly analyse the voice samples and symptoms metadata, and show that better performance can be obtained by combining them. Our study confirms that even in the most challenging scenario of in-the-wild crowdsourced data, voice is a promising signal for the pre-screening of COVID-19.

\vspace{-.1cm}
\vspace{-.3cm}
\section{Methodology}
\label{sec:method}
\vspace{-.1cm}

This section presents a comprehensive description spanning the data acquisition, preprocessing, and tasks of interest.
We note that the data collection and study have been approved by the Ethics Committee of the Department of Computer Science and Technology at the University of Cambridge.

\vspace{-.1cm}
\subsection{Data Acquisition}
\label{sec:data}
\begin{figure}[!t]
    \centering
     \begin{subfigure}[b]{0.22\textwidth}
         \centering
      \frame{\includegraphics[trim={0cm 0cm 0cm 0cm}, clip, width=\textwidth]{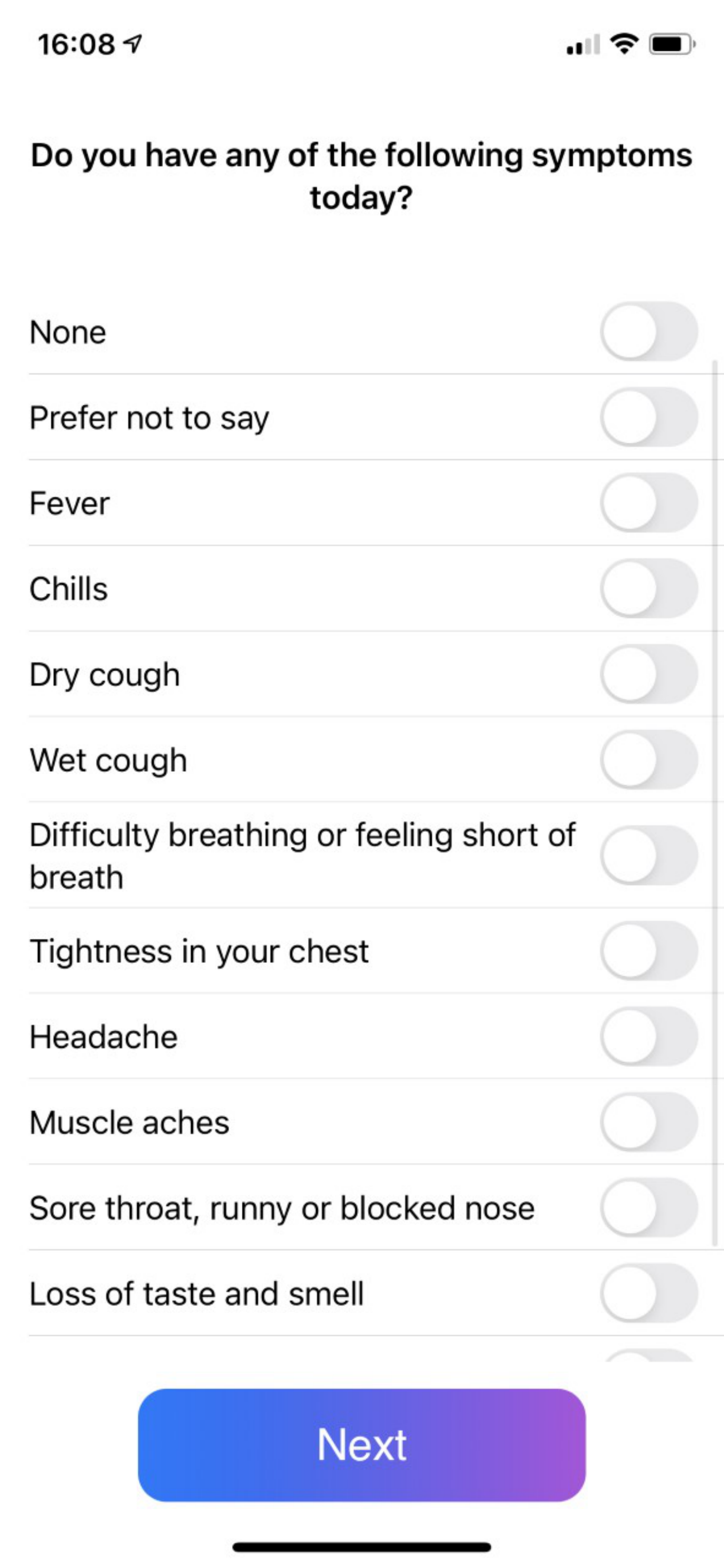}}
         \caption{Symptoms}
         \label{fig:Symptoms}
     \end{subfigure}
\hfill
     \begin{subfigure}[b]{0.22\textwidth}
         \centering
      \frame{\includegraphics[trim={0cm 0cm 0cm 0cm}, clip,width=\textwidth]{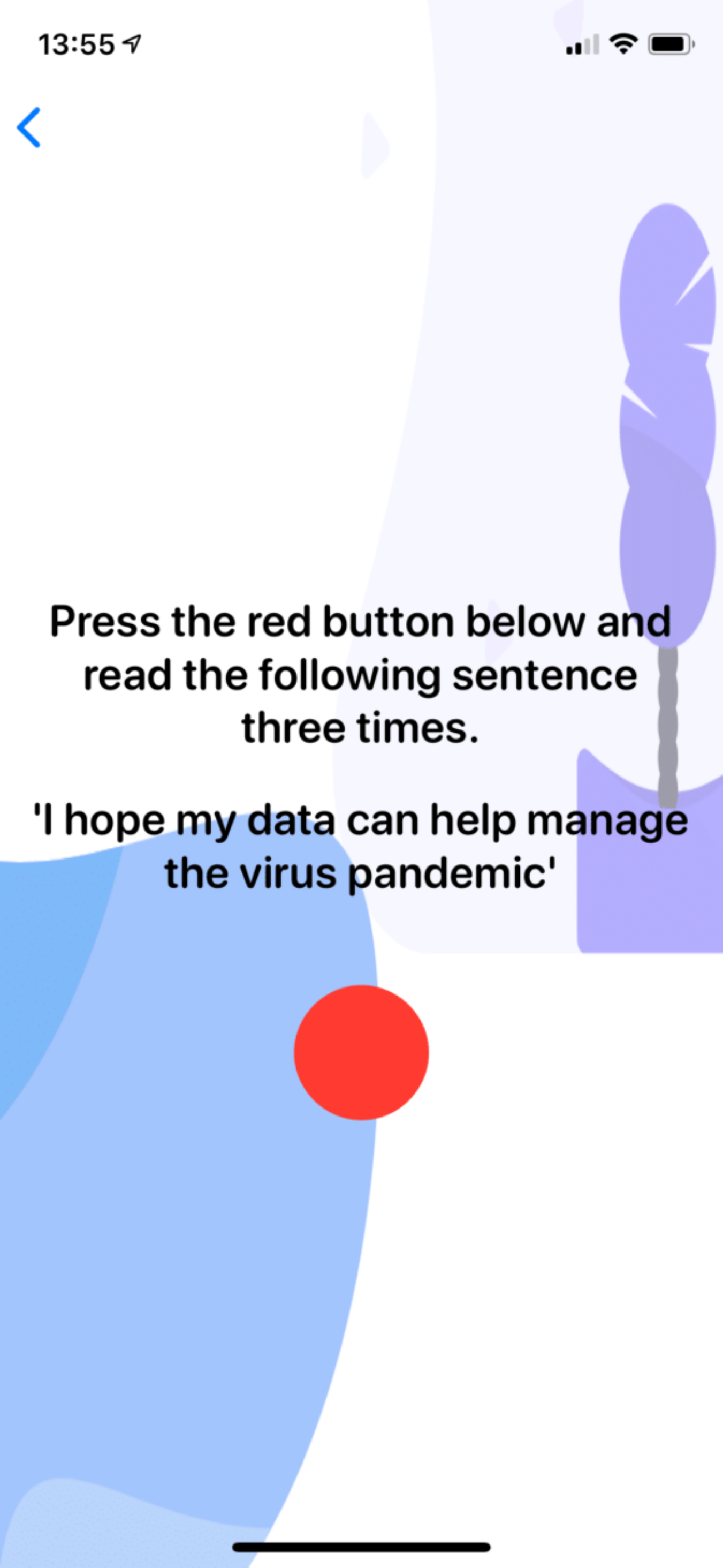}}
         \caption{Voice Recording}
         \label{fig:Recording}
     \end{subfigure}
    \caption{Screenshots of the COVID-19 Sounds App when (a) reporting symptoms, %
    and (b) recording voice samples.}
    \label{fig:app}
\vspace{-.3cm}
\end{figure}

The crowdsourced data is collected via our ``COVID-19 Sounds App~\footnote{\url{www.covid-19-sounds.org}}". It has three versions:  web-based,  Android, and iOS, with an aim to reach a high number of  users while maintaining  their anonymity. When using the app, users are asked to record and submit their breathing, coughing, and voice samples, report symptoms if any, and provide some basic demographic and medical information. Moreover, it also asks users if they have been tested positive (or negative) for the virus, and if they are in hospital. For more details of our data collection framework, the reader is referred to~\cite{Brown20-Exploring}. Fig.~\ref{fig:app} illustrates some symptom- and voice-collection screens from the iOS app.

As of 14th October 2020, data from 13722 unique users (4690 from the web app, 6334 from Android, and another 2698 from iOS) were collected. 
In this study, we explore data from two groups of participants, \ie users who declared having tested positive for COVID-19, and those who tested negative. As a consequence, data from 343 participants were selected for our analysis. In particular, 140 participants were tested positive, 199 tested negative, one transitioned from being initially positive to negative later, and another three  transitioned the other way round:  negative to positive. 

Note that in our selected subset of
data, similar to the positive participants, negative participants declared their symptoms to varying extents as well.
Likewise, there are asymptomatic positive participants who selected ``None" when asked about their symptoms. 
A comparison of the percentage occurrence of 11 symptoms (``None'' excluded) between positive and negative participants is depicted in Fig.~\ref{fig:symptomsCompare}.
It appears that loss of smell or taste is more frequently reported among positive participants than negative ones, while the differences of the percentage occurrence is rather small between positive and negative participants across other reported symptoms.

\subsection{Data Preprocessing}
\label{sec:prepro}
Recently, the potential of respiratory sounds for COVID-19 diagnosis has been explored in our previous work as well as by other researchers.
However, few research works have yet investigated the possibility of detecting Covid-19 infection from voice.
In this study, we focus on voice-based analysis for disease diagnostics, and the performed data preprocessing workflow is detailed as follows. 

First, all voice recordings from the selected users were converted to mono signals with a sampling rate of 16 kHz. Moreover, recordings that do not contain any speech signal
were discarded. Then, we 
considered applying segmentation. As mentioned previously, 
each recording consists of multiple repetitions of the given sentence by the same user, varying from one to three times. However, in our preliminary analysis, we noticed that the effect of segmentation was negligible, and that segmentation might eliminate the possible breathing differences and temporal dynamics between repetitions. For this reason, we retained only unsegmented samples for further analysis, while trimming the leading and trailing silence from each recording as in~\cite{Brown20-Exploring}. Lastly, audio normalisation was applied to each recording, aiming at eliminating the volume inconsistency across participants caused by varied devices or different distances between the mouth and the microphone.

After preprocessing, we obtained a total of 828 voice samples (326 positive and 502 negative) from 343 participants. They mostly come from the UK, Portugal, Spain, and Italy.

\subsection{Tasks}
\label{sec:task}
In this study, a series of binary classification tasks are developed. In particular, based on the dataset collected, we train models for the following clinically meaningful tasks:

\begin{itemize}
    \item \textbf{Task 1}: Distinguish individuals who have declared that they were tested positive for COVID-19, from individuals who have declared that they were tested negative for COVID-19. This is a general scenario, and we refer to this task as `\textbf{Pos. v.s. Neg.}' 
    \item \textbf{Task 2}: Distinguish individuals tested positive for COVID-19 recently in the last 14 days, from individuals tested negative for COVID-19, specifically for those with a negative test and no reported symptoms.
    We refer to this task as `\textbf{newPos. v.s. Neg. w/o sym.}' This case is set following our previous work in~\cite{Brown20-Exploring}, so as to compare the capability of voice samples with breathing and cough ones for COVID-19 diagnosis.
    \item \textbf{Task 3}: Distinguish asymptomatic individuals tested positive for COVID-19, from individuals tested negative, specifically for those healthy controls that do not have any symptom. This task is devised to investigate whether asymptomatic carriers of the disease can be identified from their voice. 
    This is of concern given the high rate of asymptomatic infection reported in the population~\cite{oran2020prevalence}.
    Therefore, identifying asymptomatic individuals may play a significant role in controlling the ongoing pandemic~\cite{oran2020prevalence}. 
    We refer to this task as `\textbf{Pos. w/o sym. v.s. Neg. w/o sym.}' 
    \item \textbf{Task 4}: Distinguish symptomatic individuals who have declared that they were tested positive for COVID-19 and have developed at least one symptom, from individuals who have declared that they were tested negative though suffering from one or more symptoms. This task is considered with an aim to understand the feasibility of voice analysis to differentiate COVID-19 from other disease such as the common-flu.
    We refer to this task as `\textbf{Pos. w/ sym. v.s. Neg. w/ sym.}' 

\end{itemize}

\begin{figure}[!t]
    \centering
    \includegraphics[width=0.484\textwidth]{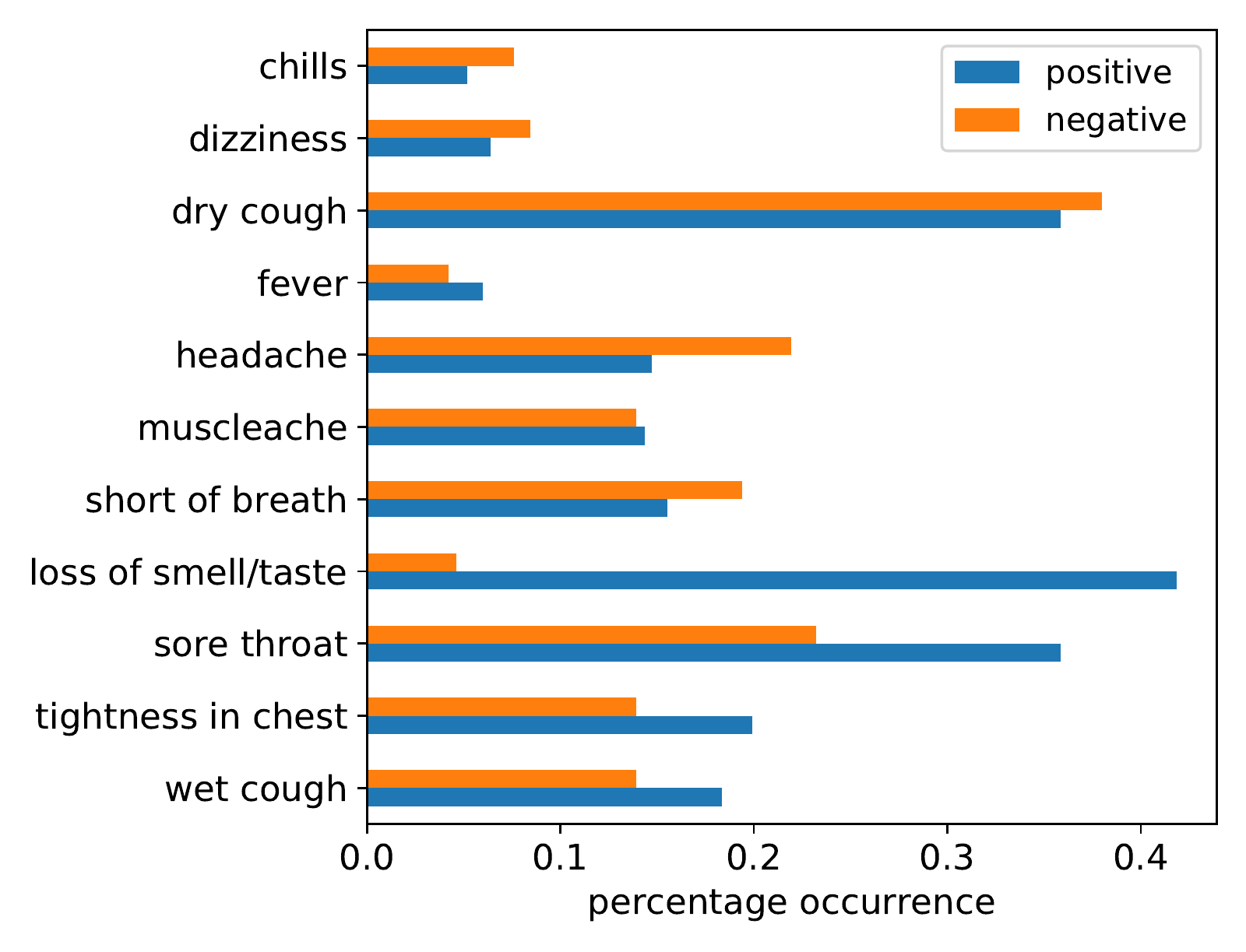}
    \caption{Comparison of the percentage of occurrence across 11 symptoms between COVID-19 positive and negative participants. 
    }
    \label{fig:symptomsCompare}
    \vspace{-.3cm}
\end{figure}

In addition to voice-based analysis, we explore the symptoms to provide complementary information. In particular, for symptomatic individuals, their voice and the symptoms are integrated as inputs to the models. 
More specifically, another three tasks are investigated:
\begin{itemize}
    \item \textbf{S$_{only}$}: Distinguish symptomatic positive individuals from symptomatic negative users by \textit{using their symptoms only}.
    \item \textbf{(V+S)$_{FF}$}: Distinguish symptomatic positive individuals from symptomatic negative users via \textit{feature-level fusion} by concatenating voice features and symptom-based features as inputs of a model.
    \item \textbf{(V+S)$_{DF}$}: Distinguish symptomatic positive individuals from symptomatic negative users via \textit{decision-level fusion} by combining the predictions from a voice-based model and another symptom-based model. In our case, the final decision will be the same as the prediction from the model with the highest probability estimate for a given instance.
\end{itemize}
 
\vspace{-.1cm}
\section{Experiments}
\label{sec:experiment}
\vspace{-.1cm}

In this section, a comprehensive evaluation is performed to investigate the performance of the tasks provided in~\ref{sec:task}. We describe the features, experiment setup, and result analysis, respectively.

\subsection{Features}
In this study, we applied an established acoustic feature set, namely the
INTERSPEECH 09 Computational Paralinguistics Challenge (\textsc{ComParE}) set~\cite{schuller2009interspeech}, extracted by an open-source openSMILE toolkit~\cite{Eyben10-openSMILE}. For each audio file, 12 functionals 
were applied on 16 frame-level descriptors and their corresponding delta coefficients, resulting in a total of 384 features.
Particularly, the 16 frame-level descriptors chosen are Zero-Crossing-Rate (ZCR), Root Mean Square (RMS) frame energy, pitch frequency (F0), Harmonics-to-Noise Ratio (HNR), and Mel-Frequency Cepstral Coefficients (MFCCs) 1-12, covering prosodic, spectral, and voice quality features~\cite{schuller2009interspeech}. For more details about these features, please refer to~\cite{schuller2009interspeech}.

Moreover, we combined the voice-based analysis with symptoms for COVID-19 diagnosis. In specific, 11 symptoms are chosen as the most common symptoms of COVID-19, as shown in Fig.~\ref{fig:symptomsCompare}.
In order to convert these symptoms into feature vectors, one-hot encoding was utilised, 
resulting in a 11-dimensional symptom-based feature vector for each sample.
Each dimension of the vector indicates the presence (1) or absence (0) of a particular symptom. 

\subsection{Experimental Setup}
Following feature extraction, we used Support Vector Machines (SVMs) with linear kernel as the classifiers for all tasks, due to its widespread usage and robust performance achieved in intelligent audio analysis~\cite{schuller2009interspeech, Han17-Prediction}. 
The complexity parameter C was set to 0.01 based on our preliminary research.
Code was implemented using the scikit-learn library in Python.

Moreover, for each task, 5-fold cross-validation was performed while the subject-independent constraint was kept, ensuring that data points from the same participant do not appear in both splits.
Further, to deal with the imbalanced data during training, data augmentation via Synthetic Minority Oversampling Technique (SMOTE)~\cite{chawla2002smote} was carried out to create synthetic observations of the minority class. 

To validate the recognition performance of the voice-based models for disease diagnosis under various scenarios, we selected the following standard evaluation metrics: sensitivity (also know as recall or true positive rate (TPR) and calculated as $TP/(TP + FN)$), specificity (also referred to as true negative rate (TNR) and calculated as $TN / (TN + FP)$), the area under the ROC curve (ROC-AUC) which measures the performance by consider both sensitivity and specificity at various probability thresholds, and the area under precision-recall curve (PR-AUC) which computes the area under the precision-recall curve.
Moreover, for each model, the mean and standard deviation across all five folds were computed separately.

\begin{table*}[!t]
\centering
 \caption{Performance in terms of sensitivity($SE$), specificity($SP$), receiver operating characteristic - area under curve ($ROC$-$AUC$), and area under precision-recall curve($PR$-$AUC$) for the \textit{voice-based} diagnosis. For each measurement, its mean and standard deviation across 5-fold cross-validation are reported.}
 \begin{threeparttable}
  \begin {tabular}{lcccccc}
  \toprule
  \textbf{Task} & \textbf{\#Pos.} & \textbf{\#Neg.} &\multicolumn{4}{c}{\textbf{mean $\pm$ std}} \\
 
  \cmidrule(l){4-7}
  & & &  \textbf{SE} & \textbf{SP} &  \textbf{ROC-AUC} & \textbf{PR-AUC} \\
  \midrule
1. Pos. v.s. Neg. &  326 & 502 & 0.62$\pm$0.15	& 0.74$\pm$0.15 & 0.74$\pm$0.08 & 0.75$\pm$0.09 \\
2. newPos. v.s. Neg. w/o sym. & 155 & 264 & 0.70$\pm$0.10 & 0.75$\pm$0.10 & 0.79$\pm$0.10 &  0.79$\pm$0.08 \\
  \midrule
3. Pos. w/o sym. v.s. Neg. w/o sym. & 72 & 264 & 0.40$\pm$0.08 & 0.74$\pm$0.08 & 0.65$\pm$0.11 &	0.72$\pm$0.09\\

4. Pos. w/ sym. v.s. Neg. w/ sym. & 254 & 238 & 0.64$\pm$0.10 & 0.75$\pm$0.10& 0.77$\pm$0.06& 0.77$\pm$0.05\\
  \bottomrule
  \end{tabular}
 \end{threeparttable}
 \label{tab:res1}
  \vspace{-.3cm}
\end{table*}

\begin{table}[!t]
\setlength{\tabcolsep}{3pt}
\centering
 \caption{Results of COVID-19 diagnosis based on voice (\textbf{V}) and symptoms (\textbf{S}), where the selected positive and negative participants report at least one symptom. Performance in terms of sensitivity($SE$), specificity($SP$), receiver operating characteristic - area under curve ($ROC$-$AUC$), and area under precision-recall curve ($PR$-$AUC$) are reported. For each measurement, its mean and standard deviation across 5-fold cross-validation are reported. Best performances are highlighted. %
 }
 \begin{threeparttable}
  \begin {tabular}{lcccc}
  \toprule
  \textbf{Method} & \multicolumn{4}{c}{\textbf{mean $\pm$ std}} \\  \cmidrule(l){2-5}
  &  \textbf{SE} & \textbf{SP} &  \textbf{ROC-AUC} & \textbf{PR-AUC} \\
  \midrule
 \textbf{V$_{only}$} & 0.64$\pm$0.10 & 0.75$\pm$0.10& 0.77$\pm$0.06& 0.77$\pm$0.05\\ 
 \textbf{S$_{only}$} & 0.62$\pm$0.02 & 0.80$\pm$0.02 & 0.73$\pm$0.04 & 0.75$\pm$0.06\\
   \midrule
 (\textbf{V} + \textbf{S})$_{FF}$  & 0.66$\pm$0.12 & 0.74$\pm$0.12 & 0.77$\pm$0.07 & 0.78$\pm$0.07\\
 (\textbf{V} + \textbf{S})$_{DF}$ & \textbf{0.68}$\pm$0.16 & \textbf{0.82}$\pm$0.16 & \textbf{0.79}$\pm$0.07 & \textbf{0.79}$\pm$0.06\\
  \bottomrule
  \end{tabular}
  \begin{tablenotes}
      \small
      \item FF: feature-level fusion, DF: decision-level fusion
    \end{tablenotes}
 \end{threeparttable}
 \label{tab:res2}
 \vspace{-.6cm}
\end{table}

\begin{figure*}[!t]
    \centering
     \begin{subfigure}[b]{0.495\textwidth}
         \centering
         \includegraphics[width=\textwidth]{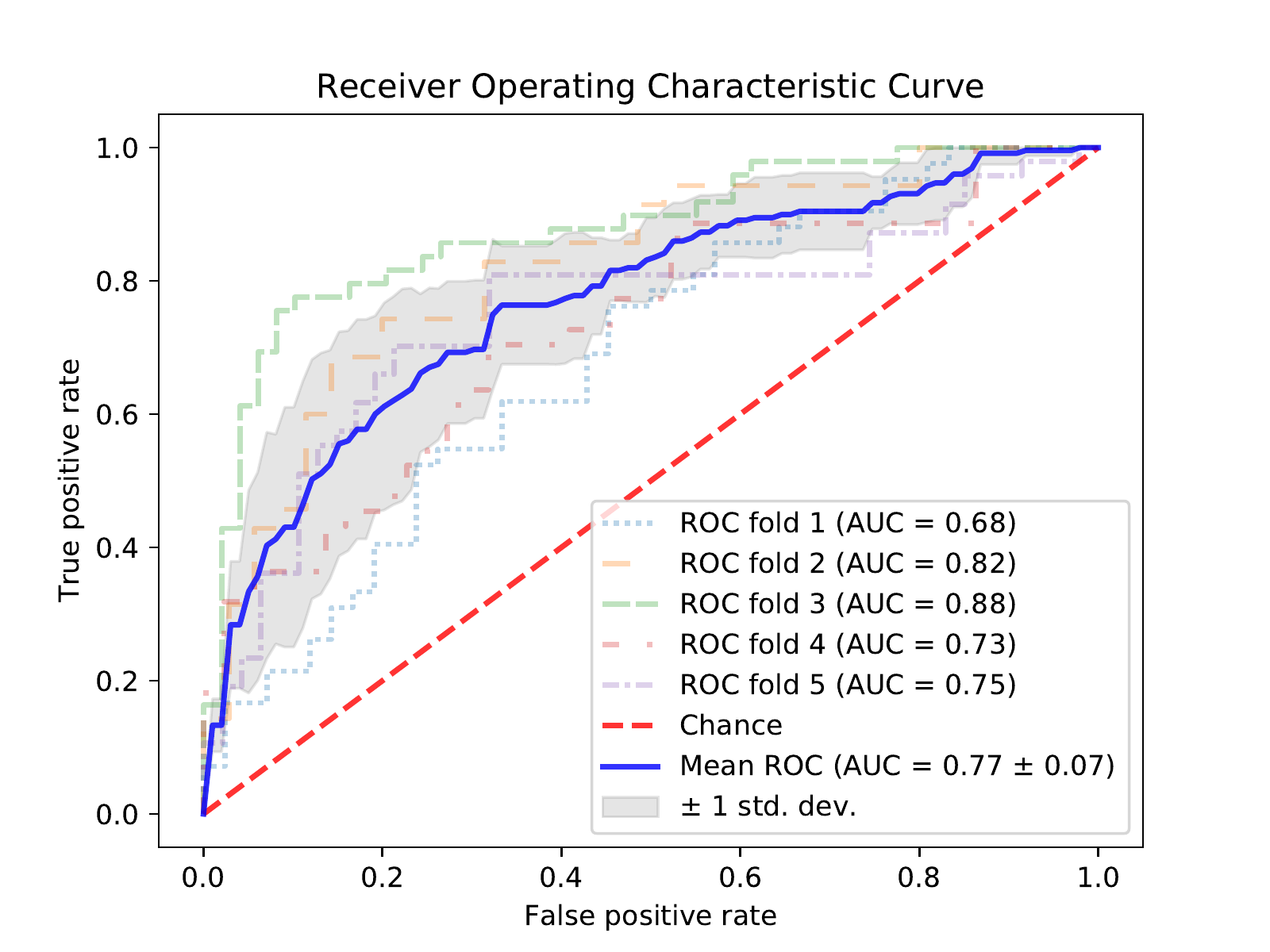}
         \caption{feature-level fusion}
         \label{fig:early}
     \end{subfigure}
\hfill
     \begin{subfigure}[b]{0.495\textwidth}
         \centering
         \includegraphics[width=\textwidth]{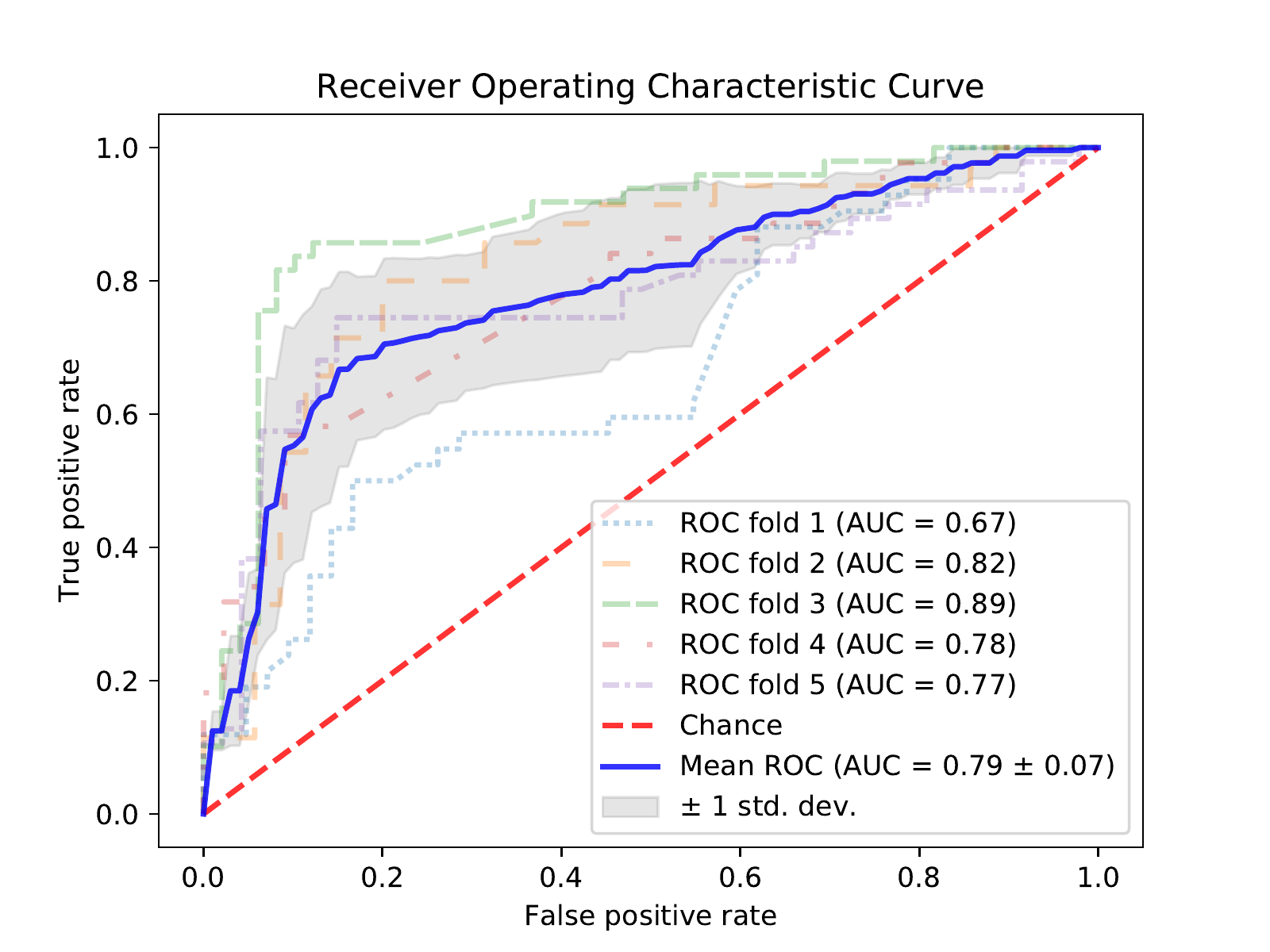}
         \caption{decision-level fusion}
         \label{fig:late}
     \end{subfigure}
    \caption{ROC curves of COVID-19 diagnosis from the combination of voice and reported symptoms, via ($a$) feature-level fusion and ($b$) decision-level fusion. In each figure, the curve of each fold under the 5-fold cross-validation is shown separately. The mean ROC curves and their variances are also given.}
    \label{fig:roc-curves}
    \vspace{-.5cm}
\end{figure*}

\subsection{Results and Discussion}
Experiment results are presented in Table~\ref{tab:res1}.
For Task 1, when distinguishing positive tested individuals from negative ones without taking their symptoms into account, the model achieves a sensitivity and specificity of $62\%$, $74\%$ respectively. Further, when distinguishing recently tested positive individuals from healthy controls without any symptoms, the ROC-AUC and PR-AUC both increase from around $75\%$ to $79\%$, while the sensitivity and specificity are improved from $62\%$ to $70\%$, and from $74\%$ to $75\%$. This indicates that voice signals have a detectable COVID19 signature. Besides, in~\cite{Brown20-Exploring}, the analysis  based on cough and breathing sounds, achieved the sensitivity of $69\%$ and ROC-AUC of $80\%$, on a different subset of users though.
It can be seen that the obtained performance from human voice is quite comparable to cough and breathing for COVID-19 diagnosis. Hence it would be interesting to analyse  all three sounds jointly to understand a comprehensive overview.

Next, when distinguishing asymptomatic patients from healthy controls (Task 3), we observe a noticeable performance decrease of the sensitivity from $70\%$ to $40\%$, indicating that a high rate of asymptomatic patients are misclassified as healthy participants. The ROC-AUC also drops from $79\%$ to $65\%$. This is in alignment with findings in a recent study~\cite{bagad2020cough}, where researchers achieved $67\%$ in ROC-AUC when identifying COVID-19 coughs from asymptomatic individuals. It implies that with the current features and model, it is difficult to identify asymptomatic patients just
from their voice.

However, when distinguishing symptomatic COVID-19 patients from non-COVID-19 controls who also developed similar symptoms (Task 4), our model achieves better performance than Task 3, attaining an AUC of $77\%$. It demonstrates the potential of exploiting voice to serve as a primary screening tool. Such a tool could rapidly identify (from symptomatic cases) the ones who might have higher priority for further clinical diagnosis.

In addition, when taking the symptoms into account, we further trained another three models. In particular, both feature-level and decision-level fusion were explored. The former concatenates audio features and the encoded symptoms as the input as a single feature matrix,
while the latter chooses the prediction with a higher probability from the two independently-trained models. Corresponding results are shown in Table~\ref{tab:res2}. When comparing the results, the best performance is achieved by decision-level fusion. It is better than each unimodal
model, attaining $79\%$ in ROC-AUC and PR-AUC, $68\%$ in sensitivity and $82\%$ in specificity. It shows the promise of combining voice and symptoms in our analysis. However, note that performance varies across folds, leading to a wide standard deviation of our models. This can also be seen from the ROC curves displayed in Fig.~\ref{fig:roc-curves}. It is believed that with more training data, it can be alleviated, as shown in our previous work~\cite{Brown20-Exploring}.

\vspace{-.1cm}
\section{Conclusions and Future Work}
\label{sec:conclusion}
\vspace{-.1cm}
In this paper, voice-based models are proposed to discriminate COVID-19 positive cases from healthy controls. The effectiveness of our models are evaluated on a crowdsourced 
dataset, and highlights the great potential of developing an early-stage screening tool based on voice signals for disease diagnosis. In addition to voice analysis, this work further explores fusion strategies to combine voice and reported symptoms which yield encouraging results.

For future work, we plan to incorporate other sounds 
such as breathing and coughing alongside voice. In addition, we will investigate the impact
of the disease on voice by analysing the correlation of voice characteristics before and after the infection. Furthermore, our data collection is ongoing, and we will improve the robustness of our models by training on a larger pool of users.

\balance

\bibliographystyle{IEEEbib}
\bibliography{refs}

\end{document}